\documentclass[prb,superscriptaddress,twocolumn]{revtex4}

\usepackage{amsmath}    
\usepackage{graphicx}   
\usepackage{hyperref}   
\usepackage{color}      
\usepackage{epstopdf}
\usepackage[caption=false]{subfig}
\begin{document}

\title{Vortex states in mesoscopic three-band superconductors}

\author{San Gillis}
\affiliation{Departement Fysica, Universiteit Antwerpen, Groenenborgerlaan 171, B-2020 Antwerpen, Belgium}
\author{Juha J\"{a}ykk\"{a}}
\affiliation{Nordita, KTH Royal Institute of Technology and Stockholm University Roslagtstullsbacken 23, SE-106 91 Stockholm, Sweden}
\author{Milorad V. Milo\v{s}evi\'{c}}
\affiliation{Departement Fysica, Universiteit Antwerpen, Groenenborgerlaan 171, B-2020 Antwerpen, Belgium}
\date{\today}

\newcommand{\todo}[1]{\textcolor{red}{(#1)}}
\newcommand{\question}[1]{\textcolor{blue}{(#1)}}

\graphicspath{{figures/}}

\begin{abstract}
Using multi-component Ginzburg-Landau simulations, we show a
plethora of vortex states possible in mesoscopic three-band
superconductors. We find that mesoscopic confinement stabilizes
chiral states, with non-trivial phase differences between the
band-condensates, as the ground state of the system. As a
consequence, we report the broken-symmetry vortex states, the chiral
states where vortex cores in different band condensates do not
coincide (split-core vortices), as well as fractional vortex states
with broken time-reversal symmetry.
\end{abstract}

\maketitle

\section{Introduction}
\label{sec:introduction}

$\text{MgB}_2$ was discovered to be superconducting in 2001
\cite{Akimitsu:2001}, as a first two-gap superconductor, with the
highest critical temperature $T_{c} = 39K$ of all binary compound
metallic superconductors known today. Comparing with other known
electron-phonon mediated superconductors, this $T_c$ is
exceptionally high, and is thought to be due to the interaction
between the two superconducting gaps. These gaps have been
experimentally measured using scanning tunneling spectroscopy
\cite{Giubileo2001,Iavarone:2002}, point contact spectroscopy
\cite{Szabo2001,Schmidt2002} and heat capacity experiments
\cite{Bouquet2001,Yang2001}. Although multiband superconductors have
been theoretically proposed over fifty years ago
\cite{Suhl1959,Moskalenko1959}, it was only after this discovery
that there was renewed theoretical and experimental interest in
multiband superconductivity.

Two-band superconductors can exhibit new phenomena not present in
conventional single band superconductors. It was theoretically
predicted that there could exist vortex states with non monotonic
inter-vortex interactions, namely short range repulsion and long
range attraction due to competing length scales of the two bands
\cite{Babaev2005,Chaves2011,Komendova2011,Shanenko2011,Lin2011},
resulting in unusual vortex patterns \cite{Moshchalkov:2009,
Nishio:2010}. Not only is the interaction between vortices markedly
different from the single band case, but it is also possible to
stabilize fractional vortices in multiband superconductors
\cite{Tanaka2001,Babaev2002,Babaev2009}. Another new phenomenon that
occurs in two-band superconductors is that of hidden criticality
\cite{Komendova2012a}. When two gaps are weakly coupled, the
coherence length of the weaker band shows a pronounced peak close to
a hidden critical temperature, in stark contrast with the monotonic
behavior of the coherence length as a function of temperature in
single band superconductors.

Related to length-scales, it is well known that confined
superconductivity on the mesoscopic scale brings forth interesting
behavior, such as enhancement of critical magnetic field
\cite{Baelus:2002,Schweigert1999} or the Little-Parks effect
\cite{Little1962, Bruyndoncx1999}. Additionally, the boundary can
impose its symmetry on the vortex matter, enabling e.g. formation of
giant vortices \cite{Kanda2004,Milosevic2009,Xu2011,
Golubovic2005,Baelus:2002,schweigert}. For two gap superconductors
it was shown that non composite and fractional vortex states can be
thermodynamically stabilized by the mesoscopic boundary
\cite{Geurts:2010, Pereira:2011, Silaev2011, Pina2012, Col2005,
Chibotaru2010, Chibotaru2007}, due to different interaction of two
band-condensates with the mesoscopic confinement.

The recent discovery of iron-based superconductors
\cite{Kamihara2008,Chen2008,Chen2008a} has further increased
interest in multiband superconductors, since these materials
typically have more than two coupled superconducting bands
\cite{Lu2009,Kuroki2008}. The coupling can impose a phase difference
between the different components of the superconducting order
parameter. This allows inherently new physics not present in single
band or two band superconductors, due to frustration between the
phase locking tendencies, leading to states with spontaneously
broken time reversal symmetry (BTRS) \cite{Garaud:2011, Garaud:2013,
Stanev2010, Orlova2013}. These states allow magnetic flux-carrying
topological solitons, with distinct magnetic signatures, which
should be observable in experiments.

In Refs. \cite{Garaud:2011, Garaud:2013} the topological solitons
with broken time reversal symmetry were considered in bulk
superconducting samples. In this paper, we study the influence of
the mesoscopic confinement on the BTRS states. We look at a system
with one strong superconducting band, which couples to two other
bands which are only superconducting due to the interaction with the
first band (the usual situation for temperatures close to $T_c$).
The coupling between bands is of the form $(+ + -)$, meaning that
the order parameters of second and third band prefer to have a phase
difference of $\pi$ while trying to attain the same phase as the
first band. It is clear that this can lead to frustration, and
non-trivial phase differences between the bands. As we will show,
such chiral superconducting state is indeed stable, and becomes the
ground state in mesoscopic samples. Moreover, the mesoscopic
boundary interacts with the phase domain walls in the sample. Those
domain walls energetically favor splitting of vortex lines in
different bands, and their interaction with mesoscopic boundary
enables a plethora of possible states unattainable in bulk or
non-chiral mesoscopic samples - including chiral fractional
vortices, where vorticity is not equal in all band condensates in
addition to broken time-reversal symmetry. We finally offer a
classification of the observed vortex states, richer than ones found
in any other superconducting system studied to date.

The paper is organized as follows. In Sec.
\ref{sec:theoretical_formalism} we introduce the Ginzburg-Landau
model for three-component superconductors. We express the
Ginzburg-Landau coefficients in the microscopic framework to
facilitate comparison with experiment. In Sec.
\ref{sec:classification} we discuss the observed vortex states and
order them in three main classes, with several subdivisions. Our
summary and conclusions are given in Sec. \ref{sec:conclusion}.

\section{Theoretical formalism}
\label{sec:theoretical_formalism}

In this paper, we perform theoretical simulations in the framework
of the three-component Ginzburg-Landau model. The governing energy
functional takes the form:
\begin{eqnarray}
& &\mathcal{F} = \int dV \left[ \sum_{i=1}^3 \left( \alpha_i
|\Psi_i|^2 + \frac{1}{2} \beta_i |\Psi_i|^4 \right) \right.
\nonumber \\
& & \left. - \sum_{i=0}^3 \sum_{j>i}^3 \eta_{ij} |\Psi_i| |\Psi_j|
\cos(\phi_i-\phi_j)\right. \nonumber \\ & &\left. + \sum_{i=1}^3
\frac{1}{2m_i}\left| \left( -i\hbar \nabla -
\frac{2e}{c}\vec{A}\right) \Psi_i \right|^2 + \frac{(\nabla \times
\vec{A})^2}{8\pi} \right], \label{freeen}
\end{eqnarray}
where $\mathcal{F}$ is the difference in free energy between the
superconducting and normal state, $\Psi_i$ are the complex order
parameters of the band-condensates (with phase $\phi_i$), $\vec{A}$
is the vector potential, $\alpha_i$, $\beta_i$ are the
phenomenological GL coefficients, $\eta_{ij}$ denote the `Josephson'
couplings between the bands, and $i$ and $j$ are the band indices.

We next rewrite Eq. (\ref{freeen}) in a dimensionless form, by
scaling length to units of $\xi_1 = \hbar / \sqrt{-2m_1 \alpha_1}$,
the order parameters to $\Psi_{10} = -\alpha_1 / \beta_1$, the
vector potential to $A_0 = \hbar c / 2 e \xi_1$ (thus magnetic field
is scaled to $H_{c_2}^{(1)} = \hbar c / 2 e \xi_1^2$) and free energy
to $\mathcal{F}_0 = \xi_1^3 \alpha_1^2 V / \beta_1$, where $V$ is
the volume of the sample. The dimensionless energy functional reads:
\begin{eqnarray}
&&\frac{\mathcal{F}}{\mathcal{F}_0} = \frac{1}{V} \int dV \left[
\sum_{i=1}^3 \left( \frac{\alpha_i}{|\alpha_1|} |\Psi_i|^2 +
\frac{1}{2} \frac{\beta_i}{\beta_1} |\Psi_i|^4 \right)\right. \nonumber \\
& &- \sum_{i=0}^3 \sum_{j>i}^3 \frac{\eta_{ij}}{|\alpha_1|} |\Psi_i|
|\Psi_j| \cos(\phi_i-\phi_j) \nonumber
\\ & &\displaystyle \left. + \sum_{i=1}^3 \frac{m_1}{m_i} \left|
\left( \nabla - i \vec{A}\right) \Psi_i \right|^2 + \kappa_1^2
(\nabla \times \vec{A})^2 \right], \label{freeen1}
\end{eqnarray}
where $\kappa_1=(H_{c_2}^{(1)})^2 \xi_1^3 / 8 \pi \mathcal{F}_0 V$.
The free energy is then minimized numerically
in order to obtain solutions to the GL model. The fields found from
this minimization will automatically be solutions to the equations
of motion:
\begin{eqnarray}
& &\frac{\alpha_i}{|\alpha_1|} \psi_i + \frac{\beta_i}{\beta_1} |\psi_i|^2 \psi_i \nonumber \\ &&-\sum_{j \neq i} \frac{\eta_{ij}}{|\alpha_1|} \psi_j + \frac{m_1}{m_i} \left( \nabla - i \vec{A} \right)^2 \psi_i=0, \label{eq:equation_of_motion_1} \\
& &\vec{J} = \sum_{i=1}^3 \vec{J}_i = \nabla \times \nabla \times \vec{A} \nonumber \\
&&= \sum_{i=1}^3 \frac{m_1}{2 m_i \kappa_1^2} \left( i \left(
\bar{\psi}_i \nabla \psi_i - \psi_i \nabla \bar{\psi}_i \right) +
|\psi_i|^2 \vec{A} \right). \label{eq:equation_of_motion_2}
\end{eqnarray}

To closer relate our results to known superconducting materials, we
express the GL coefficients in terms of microscopic parameters,
following Ref. \cite{Orlova2013}, as
\begin{align}
\alpha_i =& N(0) \gamma_{ii} - N_i(0) \mathcal{A} - N_i(0) \tau, \\
\beta_i =& N_i(0) \frac{7 \zeta(3)}{8\pi^2 T_c}, \\
\mathcal{K}_i =& \frac{\hbar^2}{2m_i} = \xi_i^2 |\alpha_i| = \frac{\beta_i}{6} \hbar^2 v_i^2, \\
\eta_{ij} =& N(0)\gamma_{ij},
\end{align}
where $N(0) = \sum_{i=1}^3 N_i(0)$ is the total density of states,
$\mathcal{A} = \ln \left( \left(2e^{\Gamma} \hbar \omega_D \right) /
\left(\pi T_c \right) \right)$ with $\Gamma$ the Euler constant and
$\omega_D$ the Debye energy, and $\tau = \ln \left( T / T_c
\right)$. $v_i$ are the band dependent Fermi velocities, and
$\gamma_{ij}$ denote the elements of the inverted interaction
matrix.

Considering that microscopic parameters of relevant materials are
not yet known with certainty, in what follows we will choose a set
of Ginzburg-Landau parameters, and just note that it is possible to
integrate real microscopic parameters in the study:
\begin{align}
\frac{\alpha_i}{|\alpha_1|} =& \left( -1,\frac{2}{3}, \frac{2}{3} \right), \nonumber \\
\frac{\beta_i}{\beta_1} =& \left( 1, 1, 1 \right), \nonumber \\
\frac{\eta_{ij}}{|\alpha_1|} =& \left( \frac{2}{3}, \frac{2}{3}, -2 \right), \nonumber \\
\frac{m_1}{m_i} =& \left( 1, 1, 1 \right). \nonumber
\end{align}
With this choice of parameters, we consider three bands with same
parameters except for the elements in the interaction matrix, i.e.
only the coupling constants between the bands will differ, as well
as the respective critical temperatures of the bands. Such choice
will enable us to more easily differentiate the effects of chirality
from the other forms of competition between the band condensates.

\begin{figure}
\includegraphics[width=3.3in]{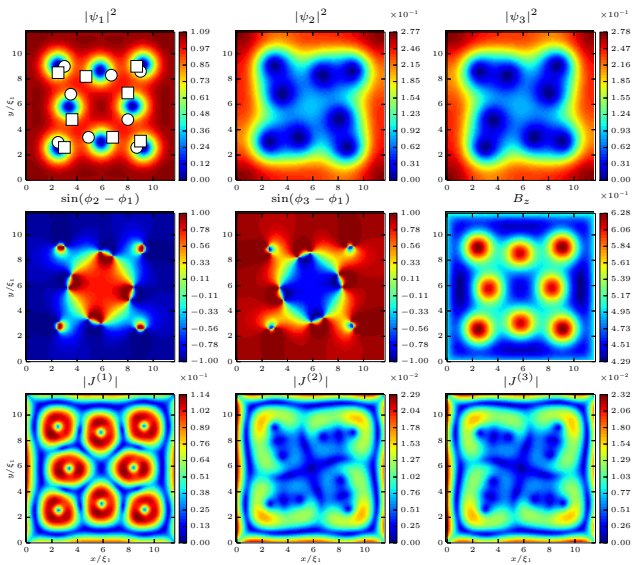}
\caption{\label{fig:start_state} One of the four initial states used
in the simulations, created by shrinking the known ($N=8$, see Ref.
\cite{Garaud:2011}) soliton solution for bulk three-band
superconductors to a mesoscopic size. Different panels show: the
Cooper-pair density of different band-condensates
($\psi_{i=1,2,3}$), the phase difference between the bands, spatial
distribution of the total magnetic field $B_z$, and magnitude of the
three components to the current ($J_{i=1,2,3}$), each stemming from
a different band-condensate. In the Cooper-pair density plot for the
first condensate (top-left panel), the locations of vortices in the
second and the third condensate are shown by white dots and squares
respectively.}
\end{figure}

In the numerical approach, we search for as many solutions as
possible to the equations of motion Eq.
(\ref{eq:equation_of_motion_1}) and (\ref{eq:equation_of_motion_2}).
We do this by starting from four different initial configurations:
(i) the Meissner state with no phase difference between the
condensates, (ii) the Meissner state with phase difference between
the condensates, (iii) a field-cooled condition, i.e. weak and
fluctuating order parameter in each band, and (iv) a state found by
recreating the $N=8$ soliton solution from Ref. \cite{Garaud:2011}
and shrinking this state to a size of $12 \xi_1$ by $12 \xi_1$. The
exemplary result of latter operation is shown in Fig.
\ref{fig:start_state}. From each of the initial states, we sweep the
external magnetic field up and down. At certain values of the
magnetic field, the system will jump to a new state with different
vorticity. These states are saved, and from each new state, we do a
new sweep of the magnetic field in both directions to uncover other
possible states.

\subsection{Numerical scheme}

At every value of the magnetic field in a sweep, we re-relax the
free energy in order to find a solution to the equations of motion
at that magnetic field. We do this by using the standard
link-variable discretisation scheme with one-point forward
differences to discretise the energy functional $\mathcal{F}$ on a
square lattice of spacing $h$. The lattice is subdivided in
$N_i\times N_j$ points, and $N_i=N_j=120$ in all simulations
reported in this paper. This discretisation has been used before in
Ref. \cite{Babaev2009}, however it is now extended to include an
arbitrary number of complex order parameters and the corresponding
Josephson couplings. The applied magnetic field $\mathbf{B}$ is
implemented by giving $\mathbf{A}$ a boundary condition such that
$\nabla \times \mathbf{A} = \mathbf{B}$ on the boundary.

Given an initial configuration $(\psi_i, \mathbf{A})$ we used either
the quasi-Newton BFGS or conjugate gradient method, as implemented
by the TAO\cite{tao-user-ref} and PETSc\cite{petsc-user-ref}
parallel numerical libraries, to find a local minimum of
$\mathcal{F}$. Other parts of the program and support routines are
an adaption of previous work done in Ref. \cite{Jaykka2011}. More
details on the discretisation can be found in Refs.
\cite{Gropp1996,Milo2010}.

\section{Classification of vortex states} \label{sec:classification}

In this section we will classify in a comprehensive manner the many
states found using the recipe described in the previous section. It
is well known that even in single-band superconductors at a given
external magnetic field there are multiple states possible, one
ground state, and other meta-stable states - but all realizable in
experiment. This can be accomplished for example by increasing the
magnetic field until a new state with different vorticity emerges.
If the magnetic field is subsequently decreased again, the new
vortex state will not be destroyed immediately, thus one has found
two different vortex states at a given magnetic field. As it turns
out, the number of meta-stable states in three-band superconductors
is far larger than in any single-band counterpart. Using the
characteristic features of those states, we classify them in three
main categories.

\subsection{Conventional vortex states} \label{subsec:conventional}

Vortex matter in mesoscopic single band superconductors is well
understood (see Refs. \cite{schweigert,Baelus:2002} and citing
articles) and we will refer to similar states in three band
superconductors as `conventional vortex states'. They comprise
composite vortices, vortex configurations influenced by the geometry
of the sample, and no phase difference between the band-condensates.
\begin{figure}[ht!]
  \includegraphics{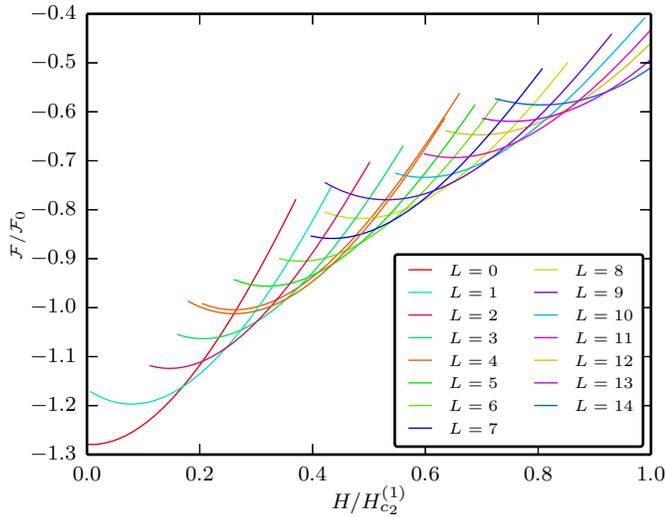}
\caption{\label{fig:conventional_energy_curves}Energy of all found
conventional (composite, not fractional, not chiral) vortex states
as a function of the applied external magnetic field. $L$ denotes
the vorticity of the state.}
\end{figure}

Fig. \ref{fig:conventional_energy_curves} shows the energy
dependence of all the conventional vortex states found during the
simulations as a function of the externally applied magnetic field.
Due to the Josephson coupling between the condensates, the second
and third condensate become superconducting. Since there is no phase
difference between the condensates, the passive bands have the same
behavior as the active band.

\begin{figure}
\subfloat[][]{\includegraphics{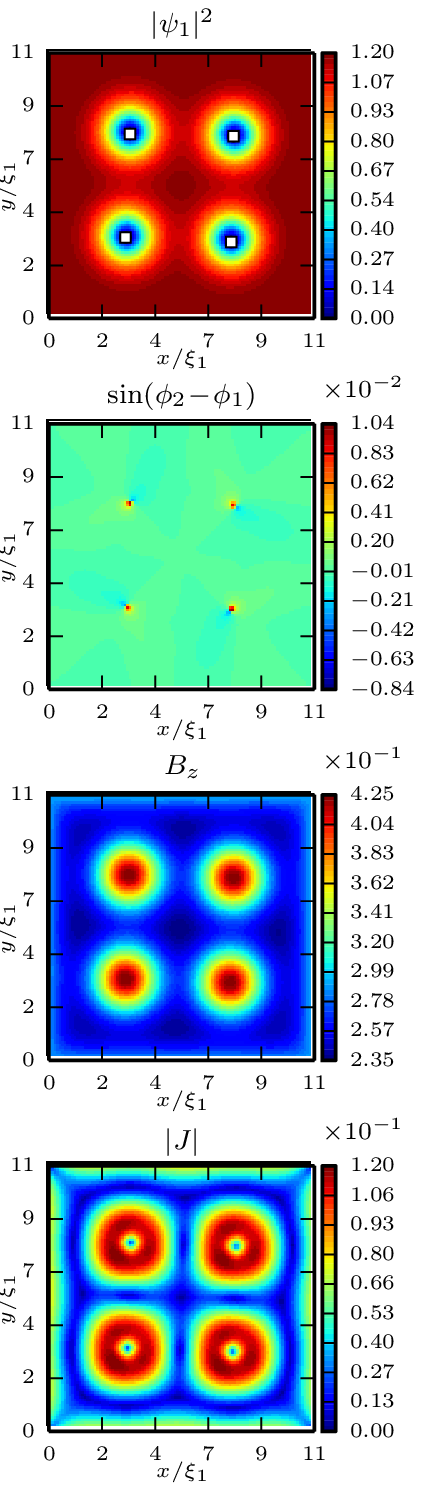}}
\subfloat[][]{\includegraphics{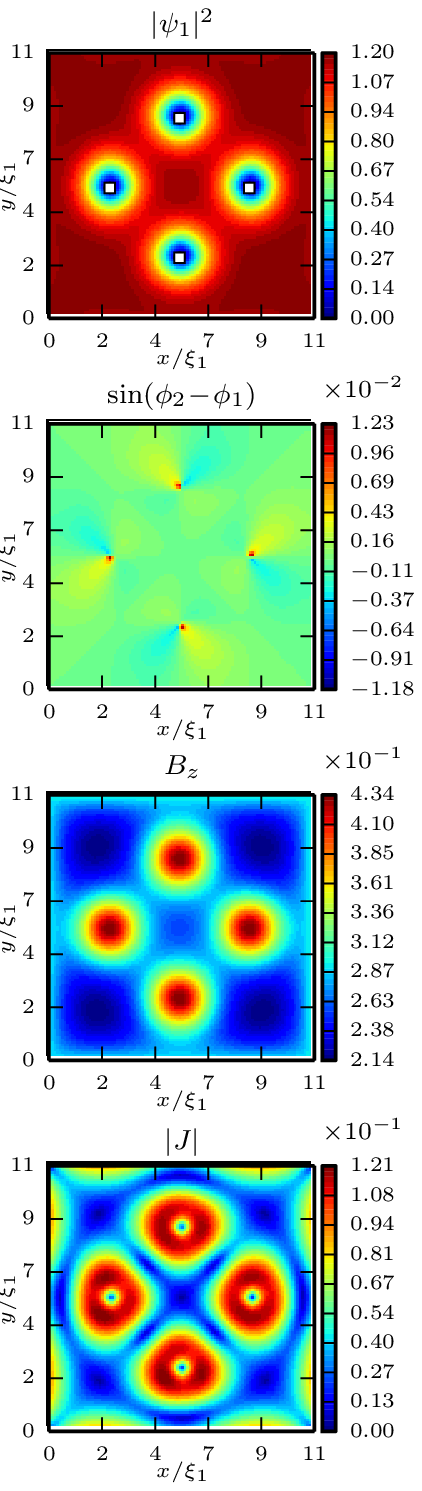}}
\caption{Conventional vortex state with vorticity $L=4$ in all bands
(a), and an alternative higher energy state (b). Panels (top to
bottom) show the Cooper-pair density of the first band with
superimposed locations of vortices in other bands, the phase
difference between the band condensates 1 and 2, the distribution of
the magnetic field, and total current in the sample.}
\label{fig:four_vortex_conventional}
\end{figure}

We note that states for each vorticity have one corresponding curve,
except for vorticity four, which has two. In Fig.
\ref{fig:four_vortex_conventional}(a) we show the lowest energy
$L=4$ state, while Fig. \ref{fig:four_vortex_conventional}(b) shows
the another stable state but with higher energy. State in Fig.
\ref{fig:four_vortex_conventional}(b) is usually not found stable in
single-band superconductors, which demonstrates subtly different
interplay of (multi-cored) vortices with screening Meissner currents
at the boundary of the multiband samples. In what follows, we show
that differences are actually very pronounced.

\subsection{Chiral vortex states} \label{subsec:chiral}

\begin{figure}
\includegraphics{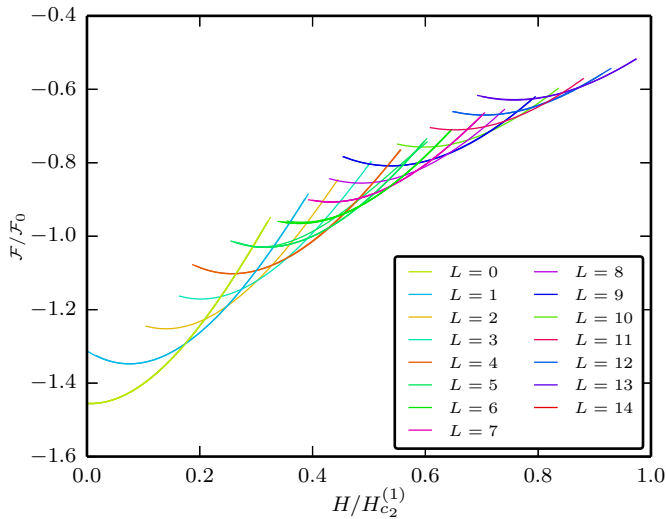}
\caption{\label{fig:chiral_red_curves}Energy of all found chiral
composite vortex states as a function of the applied external
magnetic field.}
\end{figure}

\begin{figure}
\subfloat[][]{\includegraphics{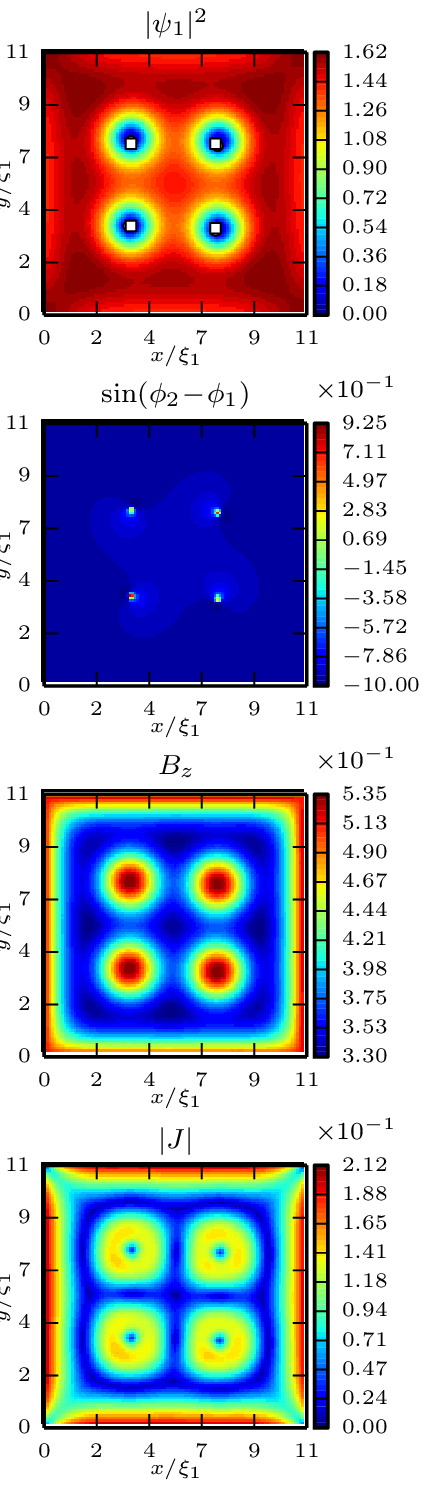}}
\subfloat[][]{\includegraphics{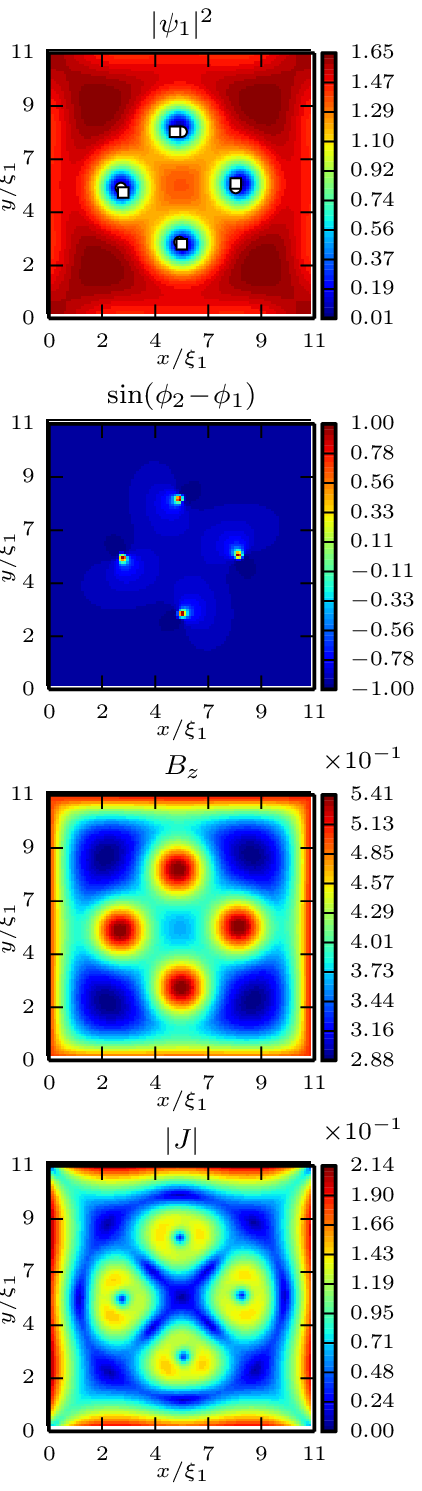}}
\caption{\label{fig:four_configuration_1} Two found configurations
of four vortices in a chiral vortex state, presented in the same
manner as Fig. \ref{fig:four_vortex_conventional}.}
\end{figure}

\begin{figure}
\subfloat[][]{\includegraphics{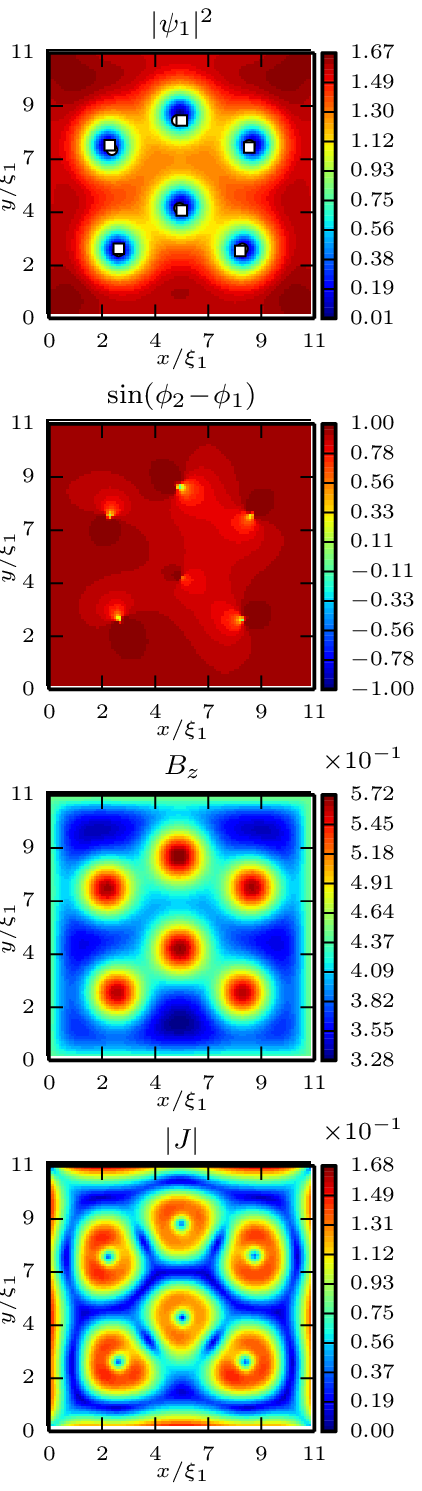}}
\subfloat[][]{\includegraphics{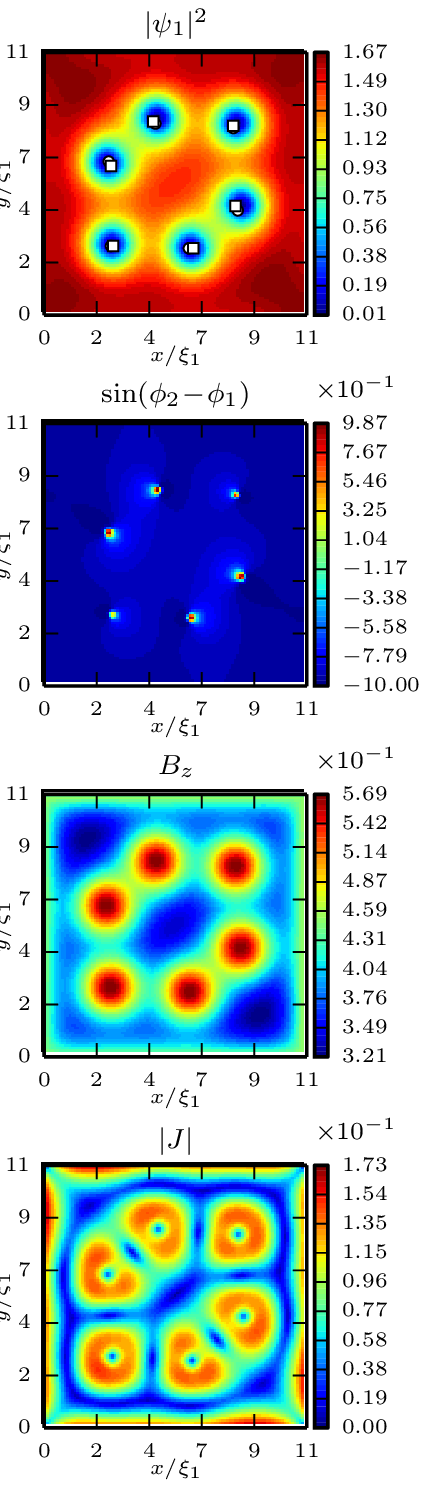}}
\caption{\label{fig:sixes_merge} Two found configurations of six
vortices in a chiral vortex state, presented in the same manner as
Fig. \ref{fig:four_configuration_1}.}
\end{figure}

It has already been shown in Refs.
\cite{Stanev2010,Garaud:2011,Orlova2013} that it is possible for
three-band superconductors to have solutions with a phase difference
between band-condensates, which we refer to as chiral solutions.
Fig. \ref{fig:chiral_red_curves} shows the energy of all found
chiral vortex states, with same vorticity in all bands. It is clear
that the basic shape of Fig. \ref{fig:chiral_red_curves} is the same
as in Fig. \ref{fig:conventional_energy_curves}, but more
importantly - the chiral states always have lower energy than the
corresponding conventional state! Since for the same parameters Ref.
\cite{Garaud:2011} reported chiral states as higher energy ones, it
is implied from our results that mesoscopic confinement enhances the
chiral states and lowers their energy compared to the conventional
states. Another important feature of chiral states is that they
exhibit much larger meta-stability, i.e. more possibilities for a
given vorticity. For example, we show in Fig.
\ref{fig:four_configuration_1} the chiral counterparts of $+$ and
$\times$ configurations from Fig.
\ref{fig:four_vortex_conventional}. However, we also find different
configurations for vorticity six for example, as shown in Fig.
\ref{fig:sixes_merge}. These are not present among the conventional
vortex states.

\begin{figure}
\includegraphics[width=3.3in]{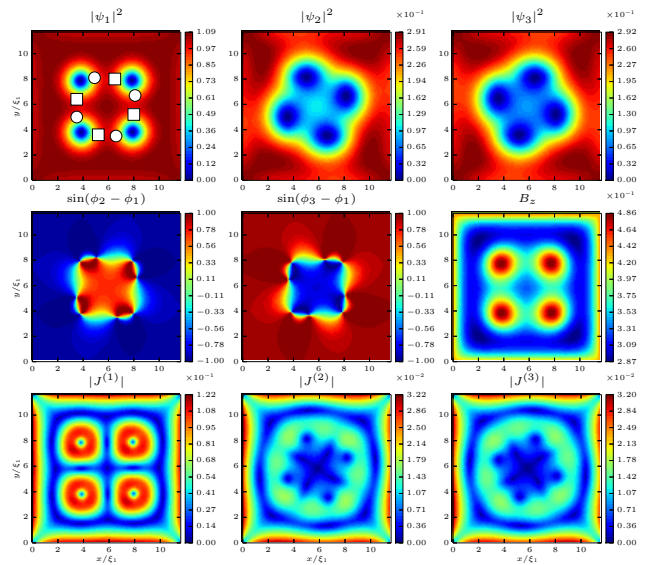}
\caption{\label{fig:four_vortex_chiral_domain_wall} Chiral $L=4$
vortex state with an internal phase domain wall, illustrated in the
same format as Fig. \ref{fig:start_state}.}
\end{figure}

\begin{figure}
\includegraphics{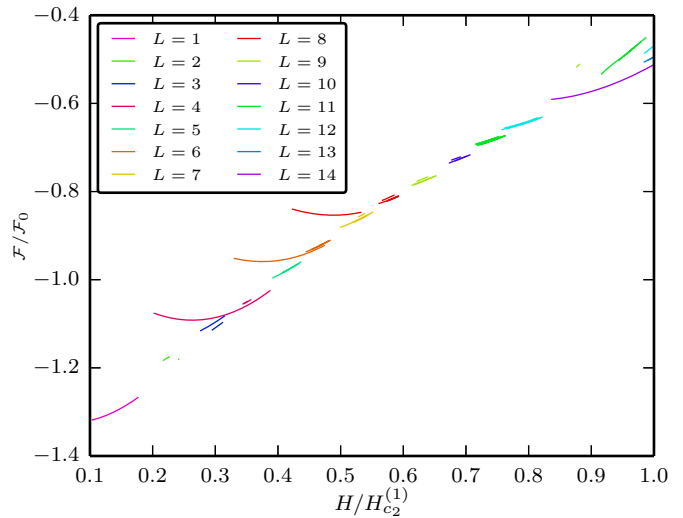}
\caption{\label{fig:chiral_domain_wall_curves} Energy curves of the
found chiral vortex states with phase domain walls, and consequently
split-core vortices.}
\end{figure}

Another factor introducing excitations into the spectrum of chiral
states with given vorticity are phase domain walls. These domain
walls were identified as a source of split-core vortex states in
bulk three-component samples in Ref. \cite{Garaud:2011}, i.e. states
in which vortex cores do not coincide in different band-condensates.
In Fig. \ref{fig:four_vortex_chiral_domain_wall} we show the
split-core $L=4$ state (c.f. Fig. \ref{fig:four_configuration_1})
where vortices in different bands minimize energy by separating
along the internal phase domain wall (connecting the vortices, see
the plot of phase difference in Fig.
\ref{fig:four_vortex_chiral_domain_wall}). However, this split-core
vortex state has higher energy than both states of Fig.
\ref{fig:four_configuration_1}. Fig.
\ref{fig:chiral_domain_wall_curves} shows the energy of all found
chiral states with domain walls, i.e. with split-core vortices. We
note three longer curves of vorticity four, six and eight. These are
states with internal domain walls as in Fig.
\ref{fig:four_vortex_chiral_domain_wall}. These states are more
stable than the other states in Fig.
\ref{fig:chiral_domain_wall_curves}, and are not present for every
vorticity - due to the competition of the vortex configuration with
the sample geometry (i.e. its C$_4$ symmetry). These states are
typically found in the simulations from the initial state shown in
Fig. \ref{fig:start_state}. States with different vorticity and an
internal domain wall were not found.

\begin{figure}
\includegraphics[width=3.3in]{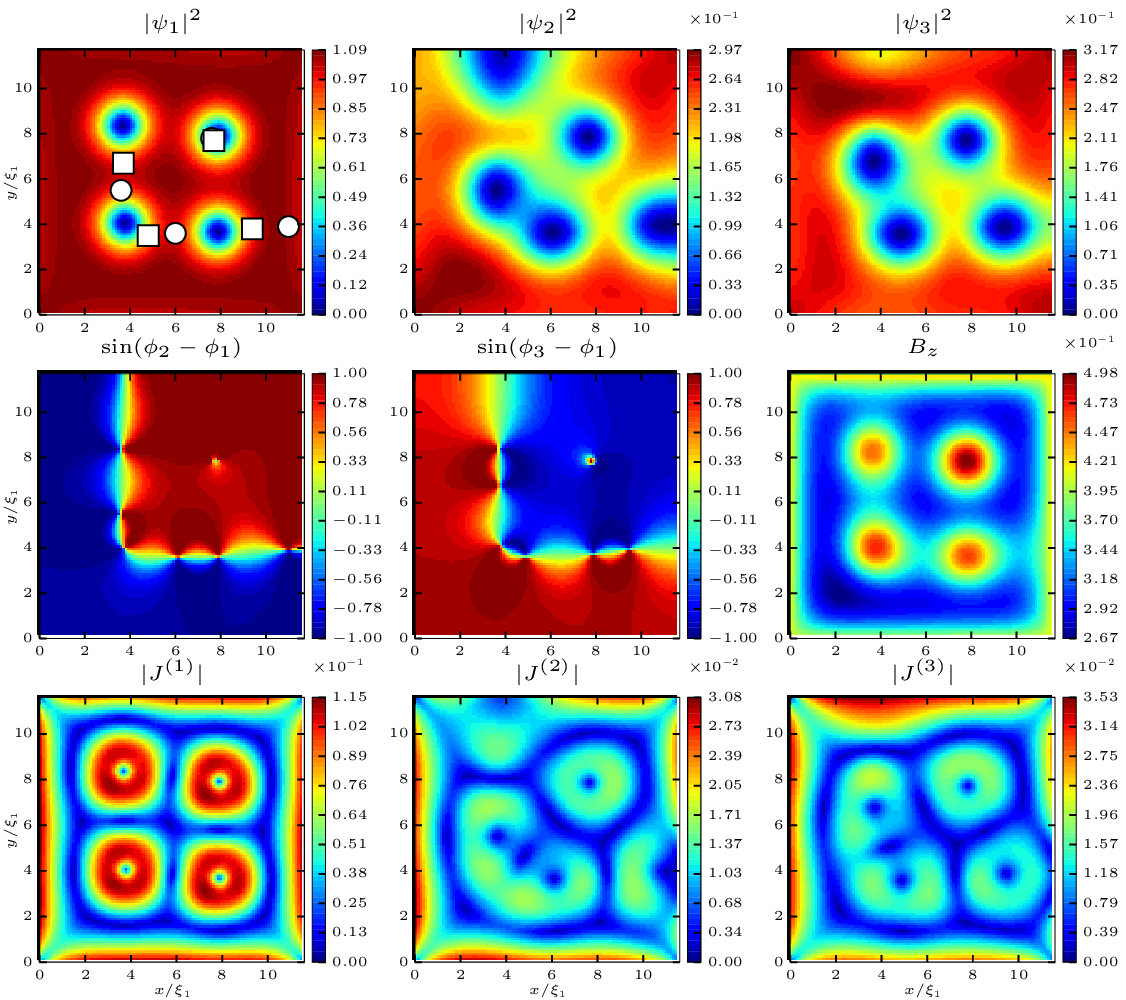}
\caption{\label{fig:four_vortex_chiral_domain_wall_2} Chiral $L=4$
vortex state with phase domain wall connecting adjacent sides of the
sample, illustrated in the same format as Fig.
\ref{fig:four_vortex_chiral_domain_wall} for facilitated comparison
of all relevant quantities.}
\end{figure}

\begin{figure}
\includegraphics[width=3.3in]{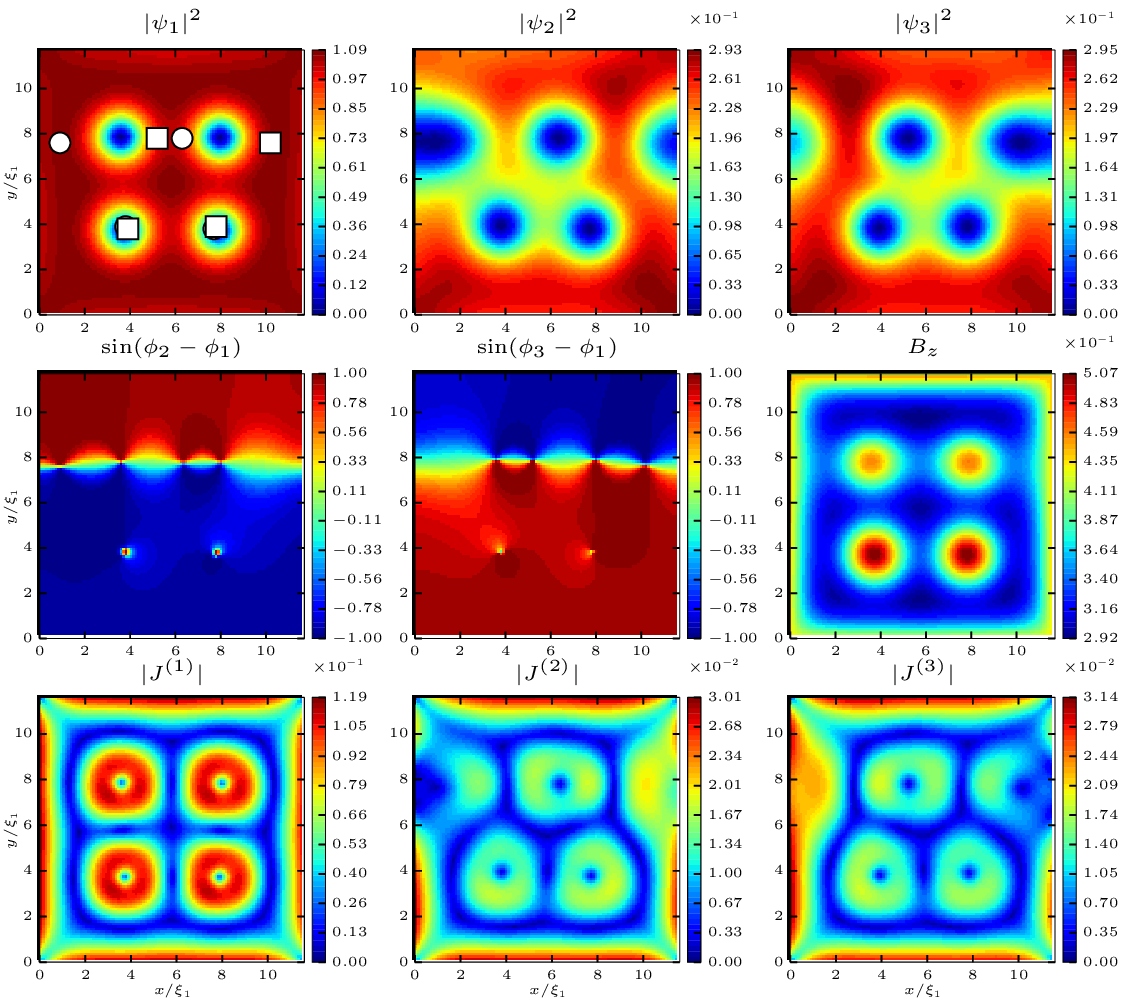}
\caption{\label{fig:four_vortex_chiral_domain_wall_3} {\it Idem.} as
Fig. \ref{fig:four_vortex_chiral_domain_wall_2}, but for a state
with phase domain spanning across the sample.}
\end{figure}

However, besides the states with an internal domain wall, we found
other states where the domain walls connect with the sample boundary
- which is a new mesoscopic effect. These states are very rich, and
can form from an arbitrary initial condition - e.g. corresponding to
a field-cooled experiment. To illustrate them, we show in Figs.
\ref{fig:four_vortex_chiral_domain_wall_2} and
\ref{fig:four_vortex_chiral_domain_wall_3} the found $L=4$ chiral
states with different geometry of the phase domain walls. In Fig.
\ref{fig:four_vortex_chiral_domain_wall_2} domain wall connects
adjacent sides of the sample, whereas in
\ref{fig:four_vortex_chiral_domain_wall_3} it spans across the
sample. Such configurations of the domain walls strongly affect the
observed vortex states, since the vortex configuration is now formed
in a threefold competition between the sample geometry, number of
vortices, and the geometry of the phase domain wall.

\begin{figure}
\includegraphics[width=3.3in]{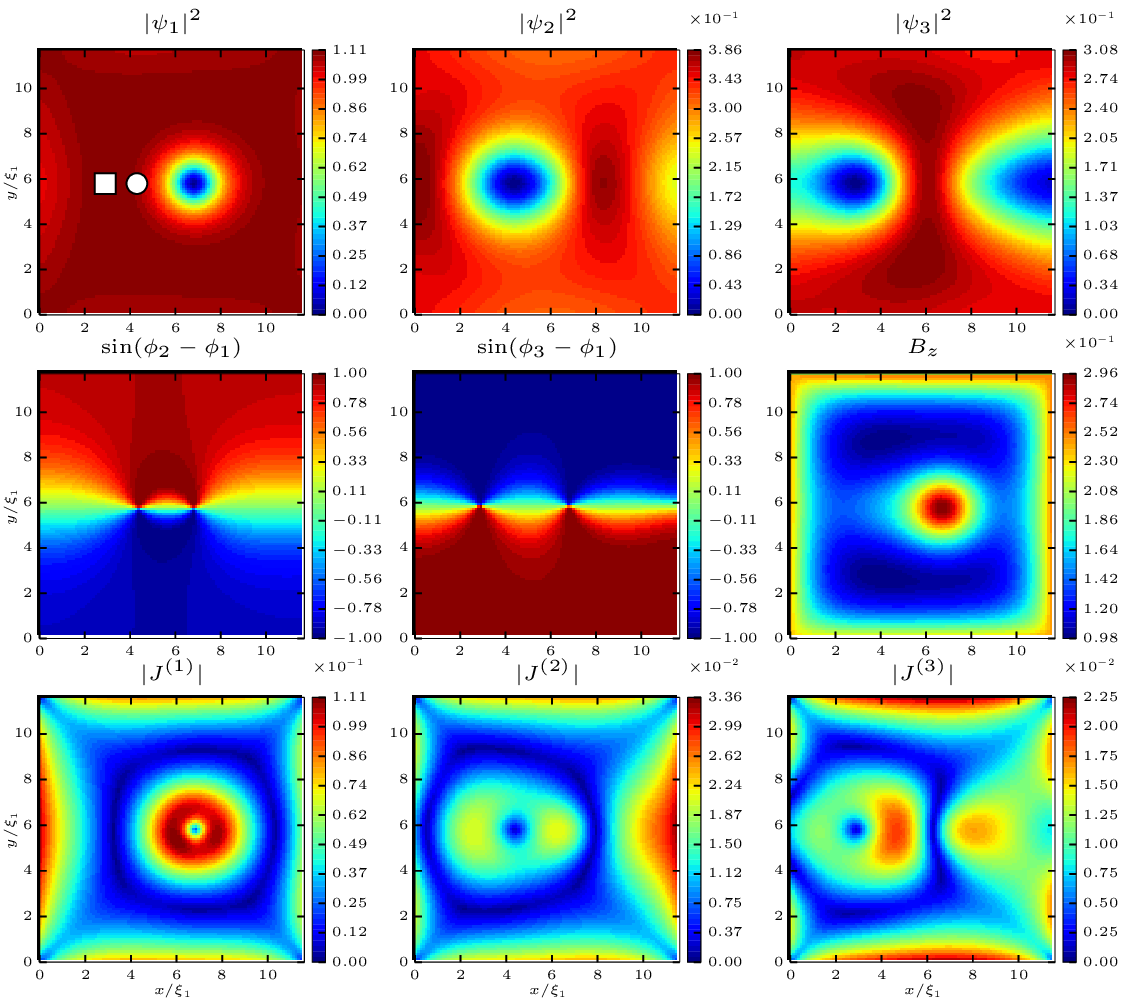}
\caption{\label{fig:broken_symmetry} Chiral vortex state for $L=1$
with notably broken spatial symmetry.}
\end{figure}

The presence of domain walls and favorable vortex splitting can
therefore result in a very pronounced symmetry breaking, as shown in
Fig. \ref{fig:broken_symmetry} for the $L=1$ state. Due to the
domain wall running across the sample, the vortices in different
bands separate along that line, and the vortex present in the first
condensate is not in the middle of the sample, which is directly
observable in e.g. scanning probe experiments. This configuration
notably breaks the fourfold symmetry. By flipping the sign of the
phase difference in both domains in Fig. \ref{fig:broken_symmetry},
the magnetic signature of the asymmetry will shift to the left
instead of to the right.

Finally, we note that although states with domain wall are more rich
and intriguing, it is actually the chiral states without domain
walls that are the ground states of this system at any given
magnetic field. This suggests that latter states will be more likely
found in experiments on chiral superconductors, but states with
domain walls remain observable in e.g. field-cooled experiments.

\subsection{Chiral fractional vortex states} \label{subsec:frustrated_fractional}

In the previous sections we showed that (1) the system has solutions
that behave as conventional vortex matter, where the passive bands
show the same behavior as the active band, and (2) that there exist
chiral solutions that actually have a lower free energy than the
conventional vortex states in the frustrated system. Excited chiral
states with phase domain walls were also possible. The latter
actually stimulates vortex splitting along the domain walls, as
discussed in Ref. \cite{Garaud:2011}. We introduced a term
`split-core vortex' as a vortex that is non composite, i.e. the
position of the vortex core in the three band-condensates is
different. On the other hand, the difference in length-scales
between the condensates\cite{Komendova2011} should further affect
the frustration, and according to Refs.
\cite{Geurts:2010,Chibotaru2010} it could also lead to fractional
vorticity, where different number of vortices is found in different
bands. The interplay of latter effects can therefore create numerous
new equilibria in the system, which we will classify by the number
of vortices in the passive bands compared to the active band for our
choice of parameters. It should be noted however, that the usual
fractional states in mesoscopic multiband superconductors, due to
only competition of length scales (as discussed in Refs.
\cite{Geurts:2010,Chibotaru2010}), are not present in our system. In
the absence of phase frustration, we have only found composite
vortices for the chosen microscopic parameters (see Fig.
\ref{fig:conventional_energy_curves}). Therefore, the fractional
vortices in the following sections are induced solely by phase
frustration and time-reversal symmetry breaking, and are therefore
called `chiral fractional vortices'.

\subsubsection{Larger vorticity in passive condensates}

\begin{figure}
\includegraphics{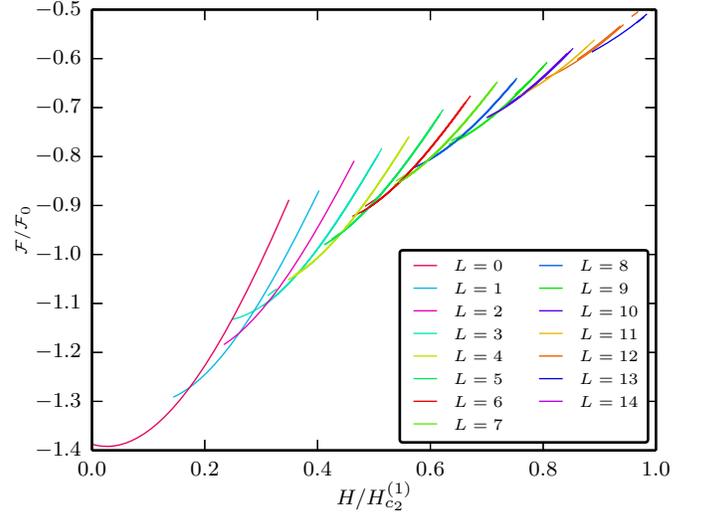}
\caption{\label{fig:green_1_walls_1_curves} Energy curves of the
chiral fractional states with one fractional vortex in the passive
bands $(L_1,L_2,L_3)=(L,L+1,L+1)$, without domains in the phase
difference between the band-condensates.}
\end{figure}

We found a multitude of fractional states with more vortices in the
passive bands compared to the active band. In Fig.
\ref{fig:green_1_walls_1_curves} we show the energy spectrum of
found states without a phase domain wall, and one additional
fractional vortex in the passive bands. The observed behavior is
fairly conventional, with exception of the less parabolic shape of
the energy curves due to the fact that fractional vortices are
easily expelled in lowered magnetic field.

\begin{figure}
\includegraphics[width=3.3in]{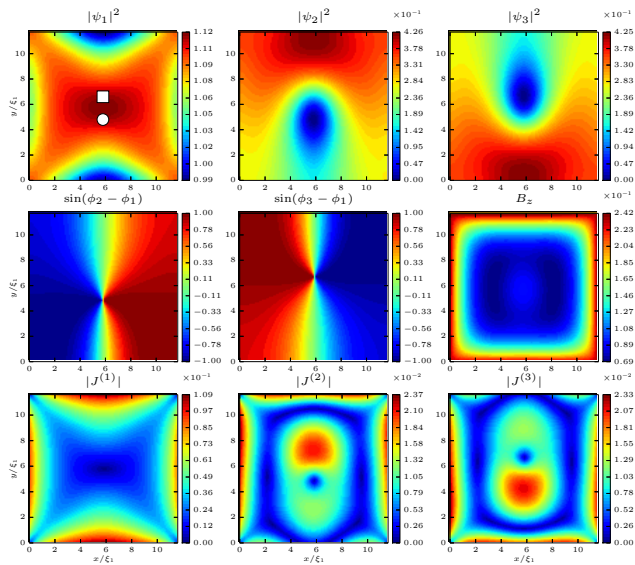}
\caption{\label{fig:green_1_vorticity_0} The chiral fractional
vortex state $(0,1,1)$.}
\end{figure}

\begin{figure}
\includegraphics[width=3.3in]{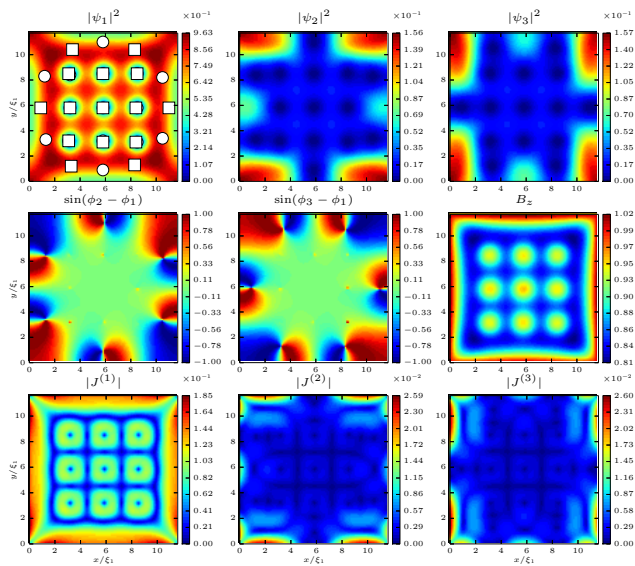}
\caption{\label{fig:green_6} The chiral fractional vortex state
$(9,15,15)$, showing the case of 6 fractional vortices aside 9
composite ones in the same sample.}
\end{figure}

In Fig. \ref{fig:green_1_vorticity_0} we show one example of a
chiral fractional state from Fig. \ref{fig:green_1_walls_1_curves}.
We find that fractional vortices avoid each other due to
time-reversal symmetry breaking, and in this particular example one
vortex can be found in passive bands, whereas none is present in the
active band. We label such state as $(L_1,L_2,L_3)=(0,1,1)$. This
state exhibits non-integer flux, spatial asymmetry of the
condensates, and stray magnetic field whose profile does not
directly show presence of any vortices. The difference in vorticity
between the bands can actually be larger than 1. In other words, it
is possible to have more than one chiral fractional vortex in the
system. In our simulations we found states with up to six (!) extra
vortices in each passive band compared to the active one, an example
of which is shown in Fig. \ref{fig:green_6} for the state
$(9,15,15)$.

\begin{figure}
\includegraphics{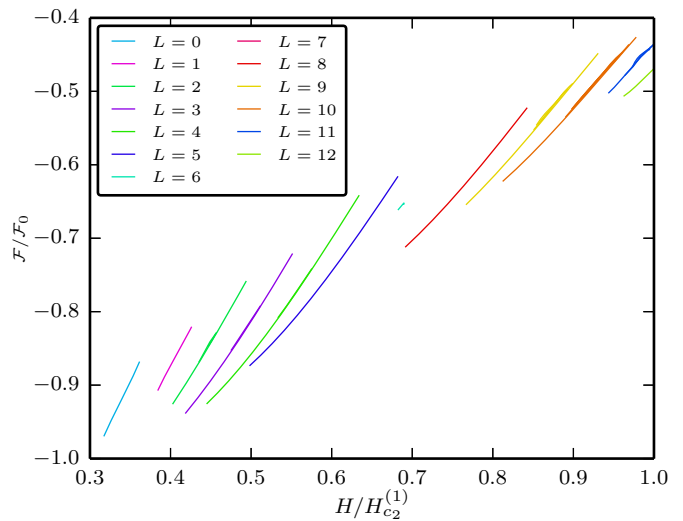}
\caption{\label{fig:green_4_curves} Energy of all found vortex
states with four chiral fractional vortices in the passive bands,
$(L,L+4,L+4)$, regardless of the phase distribution.}
\end{figure}

\begin{figure}
\subfloat[][]{\includegraphics{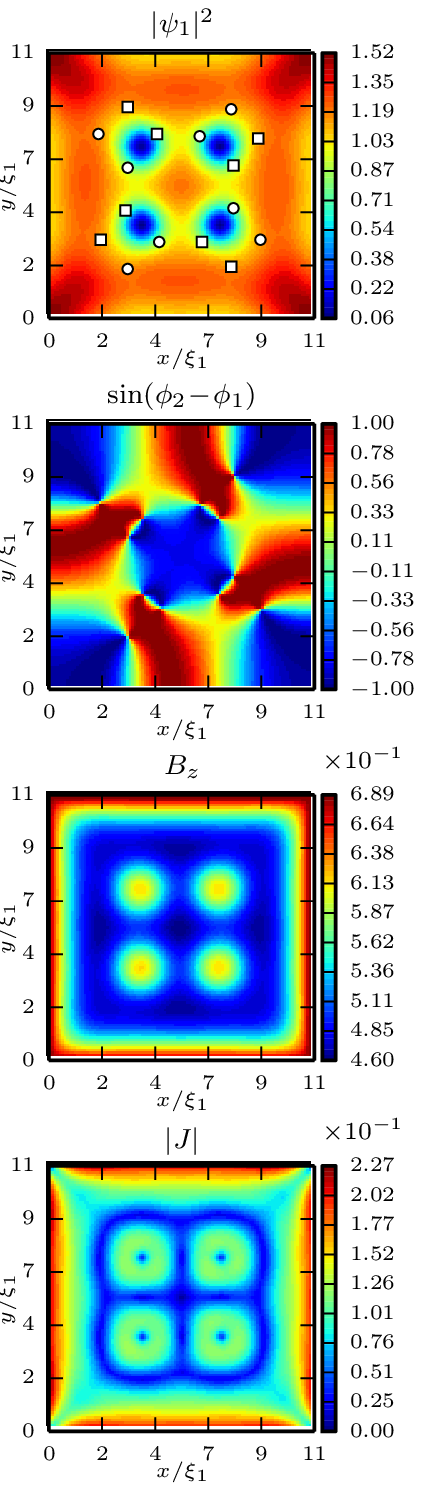}}
\subfloat[][]{\includegraphics{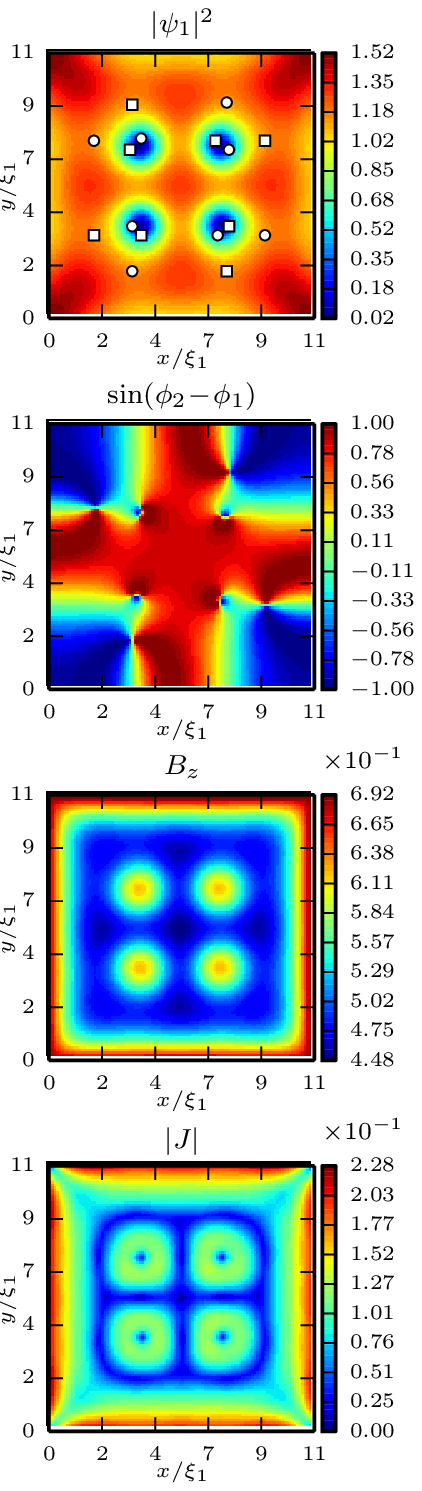}}
\caption{\label{fig:green_4_four_internal_domain} Vortex state
$(4,8,8)$, with four split-core vortices in all bands, and four
chiral fractional vortices in the second and third bands. (a) A
higher energy state, with an internal domain wall; (b) the state
with no internal domain wall.}
\end{figure}

Finally, in Fig. \ref{fig:green_4_curves} we relax the condition on
the phase, allow for the formation of the phase domain walls, and
show the energy of all found states with four chiral fractional
vortices, i.e. all possible states $(L,L+4,L+4)$. We see that some
vorticities have two or more equilibria with a slightly different
energy. We show an example of this meta-stability, found for the
state $(4,8,8)$, in Fig. \ref{fig:green_4_four_internal_domain}. As
was the case for the chiral vortex states with phase domain walls,
the states with internal domain wall always have higher energy
compared to the state without the domain wall, but show more
pronounced fractionalisation, i.e. more separated vortex cores in
the band condensates.

\subsubsection{Lower vorticity in passive condensates}

\begin{figure}
\includegraphics{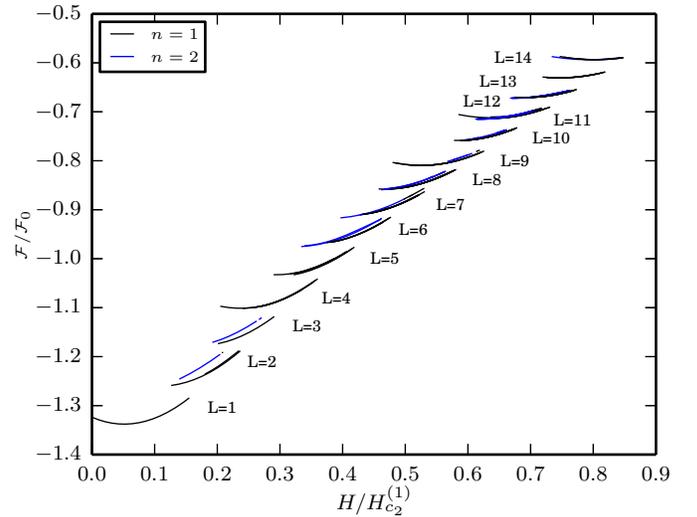}
\caption{\label{fig:blue_curves} Energy of all found states with
$n=1,2$ fewer vortices in the passive condensates than in the active
condensate, i.e. $(L+n,L,L)$ states, thus comprising $n$ fractional
vortices only in the active band.}
\end{figure}

It is also possible that the passive condensates have less vortices
than the active condensate, even though such states for our
parameters have narrower range of stability than those with more
vortices in the passive bands considered in previous subsection.
Fig. \ref{fig:blue_curves} shows the energy as a function of the
applied field for all the $(L+n,L,L)$ states, having $n=1,2$ fewer
vortices in the passive condensates. We note that for the chosen set
of parameters there are only states found for either one or two
extra vortices in the active band, indicating that fractional states
with more vortices in the active band than in passive bands are
indeed less favorable than opposite, which is of course implied by
the strong superconductivity in the active band.

\begin{figure}
\includegraphics[width=3.3in]{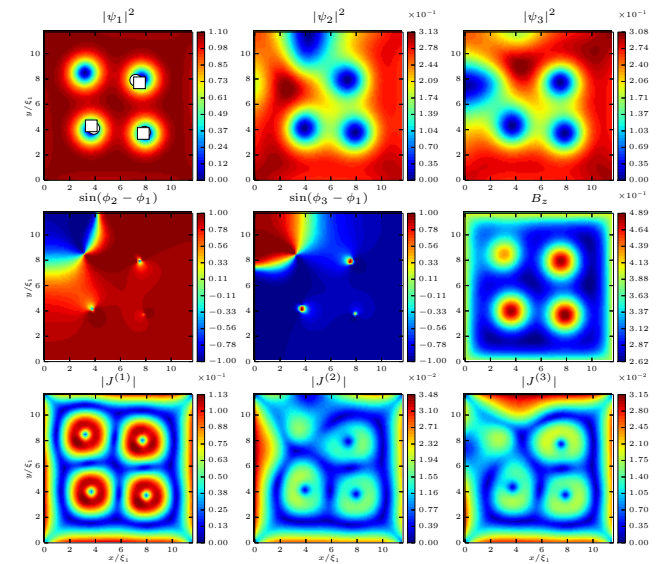}
\caption{\label{fig:blue_1_four_low_energy} Example of a chiral
state with four vortices in active band, and three vortices in the
passive bands. A phase domain wall runs through the single
fractional vortex in the first band.}
\end{figure}

\begin{figure}
\includegraphics[width=3.3in]{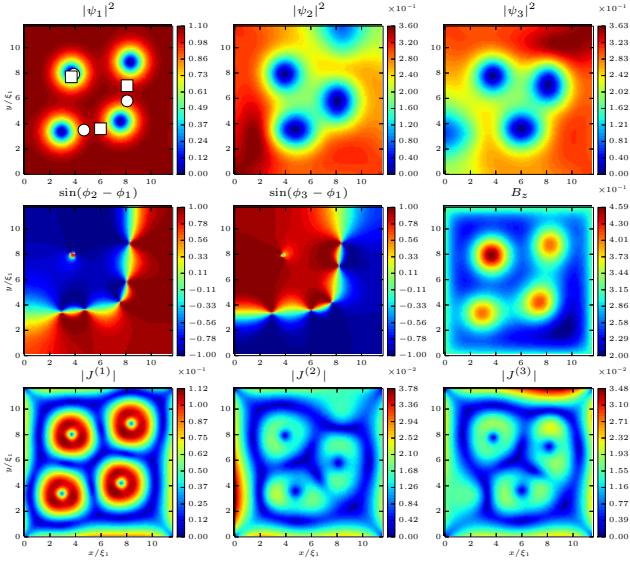}
\caption{\label{fig:blue_1_four_high_energy} Another example of a
state with same vorticity as in Fig.
\ref{fig:blue_1_four_low_energy} This state has higher energy than
the state shown in Fig. \ref{fig:blue_1_four_low_energy} due to the
different symmetry and arrangement of vortices along the domain wall.}
\end{figure}

In Fig. \ref{fig:blue_1_four_low_energy} we show an example of the
$(4,3,3)$ state with one extra vortex in the active condensate
compared to the passive condensates $(n=1)$. The frustration is
still visible in the loci of the vortex cores, although the
fractional vortex is now only present in the active band. However,
contrary to other examples of chiral fractional vortices (e.g. in
Fig. \ref{fig:green_4_four_internal_domain}) which were located in
the passive bands, the fractional vortex in the active band leaves a
clear magnetic signature in the spatial distribution of the magnetic
field compared to the three composite vortices, observable by
magnetic scanning microscopies (MFM, SHPM, etc.). It is peculiar
that in this example we found a phase domain wall running exactly
through the fractional vortex, which makes one wonder if other
possibilities for the geometry of the domain wall are stable. In
Fig. \ref{fig:blue_1_four_high_energy} we show one such possibility
for the $(4,3,3)$ state, exactly opposite to the case of Fig.
\ref{fig:blue_1_four_low_energy}. Now only one vortex is composite,
and domain wall runs through remaining three vortices of the first
band, and two in each other band, so that we seemingly have one
composite and three fractional vortices in the magnetic response of
the sample. In fact, there is only one truly fractional vortex on
the domain wall, and other two are split-core vortices.

\subsubsection{Different vorticity in passive condensates}

\begin{figure}
\includegraphics{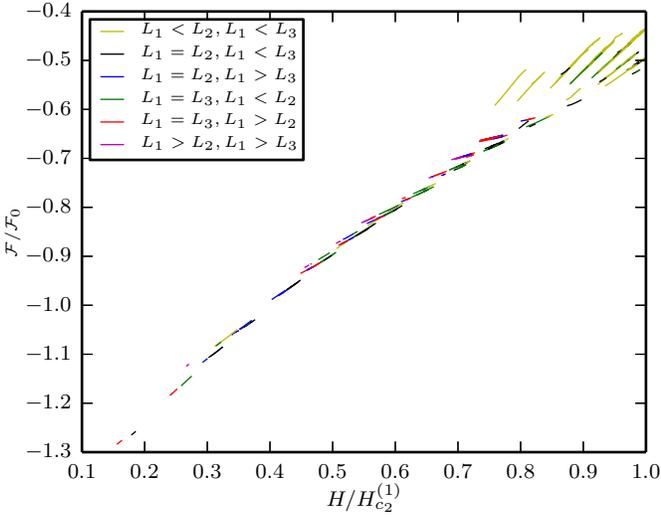}
\caption{\label{fig:uneven_curves} Energy of all found vortex states
where the number of vortices in the second and third condensates are different.}
\end{figure}

\begin{figure}
\includegraphics[width=3.3in]{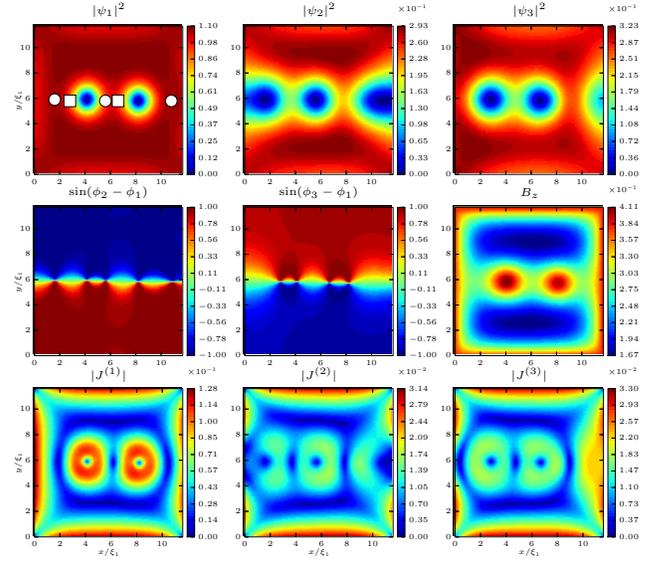}
\caption{\label{fig:uneven_two} Vortex state with three vortices in
the second condensate and two in the first and third condensate,
with an internal phase domain wall and clear symmetry breaking.}
\end{figure}

The remaining class of chiral fractional states comprises ones where
the number of vortices differs even between the two passive
condensates. Fig. \ref{fig:uneven_curves} shows the energy
dependence on the externally applied magnetic field of all
fractional states with different vorticity in the passive
condensates. An example of such state is shown in Fig.
\ref{fig:uneven_two}, for the $(2,3,2)$ case, i.e. vorticity two in
the first and third condensate, but vorticity three in the second.
These states are essentially formed in the transition between the
states discussed in the previous sections, where one chiral
fractional vortex would leave the system before the accompanying
fractions vanish as well, and are therefore much less stable than
states shown in the previous sections. The exception are the states
with larger vorticities in the passive bands, at higher applied
fields; they are seemingly more stable, but that is of pure academic
value since at such high fields and vorticities the passive bands
are extremely depleted. It is worth noting here that the rarity of
such states should be expected since the parameters of the passive
bands are taken identical. Should one investigate systems where all
the bands have significantly different parameters, one would expect
these states to become more common.

\section{Conclusion} \label{sec:conclusion}

In summary, we have studied in detail the mixed state of a
mesoscopic three-band superconductor with a frustrated choice of
Josephson couplings between the band-condensates, using the
three-component Ginzburg-Landau theory. We examined, described, and
classified all the found vortex states in a mesoscopic three-band
system (please see the interactive visualization tool of all states
in the Supplementary Material), some of which can be called {\it
conventional} - with composite vortex cores in different bands (i.e.
their cores are coaxial in three condensates), and no phase
difference between the band-condensates. As a first important
result, we showed that the ground state of the system are the {\it
chiral} states, in which phase difference is found between the band
condensates, but vortices are still composite. This is the first such
example of the chiral state as the ground state of the system at all
magnetic fields. On the other hand, a state with non-composite
(called `split-core') vortices was predicted as an excited state in
Ref. \cite{Garaud:2011} but for an existing internal phase domain
wall on which vortices are located. Indeed, we have also found such
{\it chiral split-core} vortex states in the mesoscopic system,
where the presence of domain walls not only introduces
split-core vortices, but can also lead to symmetry breaking, in more
ways than one since the domain wall can have different
configurations with respect to sample boundaries. Finally, in a
mesoscopic system, the different effect of confinement on different
condensates can stabilize states with fractional flux, i.e.
different vorticity in different bands, as is well known from the
earlier two-band studies \cite{Geurts:2010,Chibotaru2010}. In the
chiral case, vortex fractions in different bands avoid each other,
and can differ in numbers, which opens the {\it chiral fractional}
class of states. Here fractionalisation follows from the domain
walls of the phase difference between bands interacting with the
sample boundaries and favoring vortex splitting, not from different
length scales of the condensates. This plethora of distinct classes
of vortex states makes a three-band mesoscopic system an excellent
playground for further theoretical and experimental studies, where
dynamics of condensates, vortices, and phase slippage
\cite{Golib2009} must be correspondingly rich, especially since many
of the recently discovered iron-based superconductors have $3+$
overlapping bands.

\begin{acknowledgements}
This work was supported by the Flemish Science Foundation (FWO).
Critical remarks of Lucia Komendov\'{a} are gratefully acknowledged.
\end{acknowledgements}

\clearpage
\bibliographystyle{revtex}
\bibliography{bibliografie1}

\end{document}